\documentclass[12pt]{iopart}
\usepackage{cite}
\usepackage{setspace}
\usepackage{graphicx}
\usepackage{url}

\begin{document}
\title{High Precision Measurements Using High Frequency Gigahertz Signals}

\author{Aohan Jin,$^1$ Siyuan Fu,$^1$ Atsunori Sakurai,$^2$ Liang Liu,$^1$ Fredrik Edman,$^1$ Viktor \"Owall,$^1$ T\~onu Pullerits$^2$ and~Khadga J. Karki$^{2,*}$ }
\address{$^1$ Department of Electrical and Information Technology, Lund University, Ole R\"omers v\"ag 3, 22363, Lund, Sweden} 
\address{$^2$ Chemical Physics, Lund University, Getingev\"agen 60, 22241, Lund, Sweden}
\address{ Email: khadga.karki@chemphys.lu.se}

\begin{abstract}
Generalized lock-in amplifiers use digital cavities with Q-factors as high as 5$\times 10^8$. In this letter, we show that generalized lock-in amplifiers can be used to analyze microwave (giga-hertz) signals with a precision of few tens of hertz. We propose that the physical changes in the medium of propagation can be measured precisely by the ultra-high precision measurement of the signal. We provide evidence to our proposition by verifying the Newton's law of cooling by measuring the  effect of change in temperature on the phase and amplitude of the signals propagating through two calibrated cables. The technique could be used to precisely measure different physical properties of the propagation medium, for example length, resistance, etc. Real time implementation of the technique can open up new methodologies of in-situ virtual metrology in material design.
\end{abstract}

\noindent{\it Keywords\/}:Digital Cavity, Generalized lock-in amplifier, Signal generators, Microwaves, Frequency analyzers, Signal processing 

\pacs{0630Bp, 0630Ft, 0757Ty}
\submitto{\MST}
\maketitle

\section{Introduction}
Some of the most advanced techniques like digital lock-in amplifier~\cite{FREITAS_1979,MARQUES_2004,BONETTO_2005} have been designed to analyze signals up to few tens of mega-hertz frequencies. Cavity based analog techniques are still in use at high frequencies, gigahertz and higher, for high precision signal measurements. Cavities employ self interference of the signal as the mechanism of signal analysis. When a radio or microwave signal is fed into the cavity, it bounces off the walls of the cavity and interferes with itself. If twice the length of the cavity matches a period (or an integer multiple of a period) of the signal, the signal builds up in the cavity due to constructive interference, otherwise it dies out due to destructive interference. The precision with which a cavity is able to measure a signal is given by the number of periods that interfere, which in turn is given by the number of times the signal bounces off the walls of the cavity. As any physical surface, except a super conducting surface, absorbs a part of the energy from the signal that bounces off it, the number of periods that interfere is finite at any time. This together with the dielectric losses~\cite{TOMBARI_1985}  limit the finesse or the Q-factor (quality factor) of the cavity. The highest Q-factor achievable today in an analog design is about 10$^6$,~\cite{COLLIN_2001} i.e. one can measure a signal at mega-hertz frequencies with a precision of a few hertz. 

Recently, we have proposed and verified digital cavities~\cite{KARKI2013A, KARKI2013C} that mimic the functionality of  analog cavities but can reach much higher Q-factor (10$^8$ -- 10$^9$). In this letter we present different aspects of digital cavities, show that the algorithms implementing the digital cavity are suitable for ultra-high precision measurements of microwave signals digitally and discuss new kinds of metrological applications that become feasible with such techniques.

\subsection{Digital cavity} A digital cavity operates by summing up a digitized signal after each period. The period $T$ of a digital cavity is given by 
\begin{equation}\label{EQ1}
T = n \times \Delta t,
\end{equation}
 where $\Delta t$ is the digitization interval of the ADC (analog to digital converter) used and $n$ is the number of sampling points used in the digital cavity. The response $y$ of the digital cavity to a signal $x$ is given by~\cite{KARKI2013A}

\begin{equation}\label{EQ2}
y(j\cdot \Delta t;n):= \sum_{k=0}^{N_c} x(j\cdot \Delta t+k \cdot T); \quad\quad j=1,2, \cdot\cdot\cdot, n,
\end{equation} 
where $N_c$ is the cavity fold, i.e. the number of periods that are summed up in the cavity. \Eref{EQ2} captures the interference effects in a standing wave. Like in the case of an analog device, only the signals with frequencies $f$ that match the period of the cavity, i.e. $f=1/T$ (or $k f = 1/T,$ where $ k$ is an integer) `interfere' constructively in the digital cavity, while the other signals `interfere' destructively. Just as in the case of an analog cavity, the Q-factor of a digital cavity depends on the number of periods of the signal, $N_c$, that interfere in the cavity.
%the Q-factor of the digital cavity is given by $N_c$, i.e. the number of periods that interfere in the cavity. 
Unlike in the case of an analog device, the Q-factor of the digital cavity is not affected by any power loss as the information in the signal is digitized and processed in the digital domain before it is lost due to the power loss. However, there is an upper limit to the Q-factor due to the amount of memory available for storing the information. A digital cavity with an eight bit ADC running at a sampling rate of 1 GS/s (giga samples per second), and tuned to 100 MHz ($n=10$) requires a memory of about 10 MB (mega bytes) to achieve a Q-factor of 10$^6$ for continuous operation. As modern computers and DSP (digital signal processing) systems can easily accommodate GBs (giga bytes) of memory, digital cavities with Q-factors over 10$^8$ can thus be easily be achieved.    

The frequencies that are amplified by the cavity are called the resonant frequencies of the cavity. The resonant frequencies of an analog cavity can be changed by changing the length of the cavity. 
The resonant frequency of a digital cavity can be tuned by changing $n$ or $\Delta t$. It is straightforward to change $n$ in a software. But changing $\Delta t$, which allows for fine control of the resonant frequency, requires the ADCs to be clocked by a tunable clock. This specific hardware requirement is not available in all the general purpose ADC cards, hence a digital tuning technique, which is also known as the generalized lock-in amplifier,~\cite{KARKI2013C} has been implemented for ultra-high precision signal analysis using general purpose ADC cards.

\subsection{Generalized lock-in amplifier} A generalized lock-in amplifier combines the features of a traditional digital lock-in amplifier and a digital cavity to circumvent the problem of fine tuning of the resonant frequency of a digital cavity. In this method, the digital cavity is used as a line filter. The resonant frequency of the cavity is fixed. Instead the signal of interest is multiplied by a cosine (or sine) function to up/down-shift the frequency of the signal to the resonant frequency of the cavity. For example, if the signal is at 2 MHz and the resonance of the digital cavity is set to 10 MHz, then one can multiply the signal with a cosine function, $\cos(2 \pi f_{shift})$ with $f_{shift}=8 $ MHz, to upshift the signal to 10 MHz, and filter it out from the noise by using the digital cavity. Though a generalized lock-in amplifier, has some limitations, 
e.g. the numerical precision of the computer generated trigonometric functions limit the Q-factor to $5\times 10^8$,~\cite{KARKI2013C} it can be used with existing ADC cards without any additional features for digital signal processing. In the following paragraphs we present the results of extreme precision signal analysis of microwave signals by using the generalized lock-in amplifier.

\section{Experimental setup}
\begin{figure}[!t]
\centering
\includegraphics[width=3.5in]{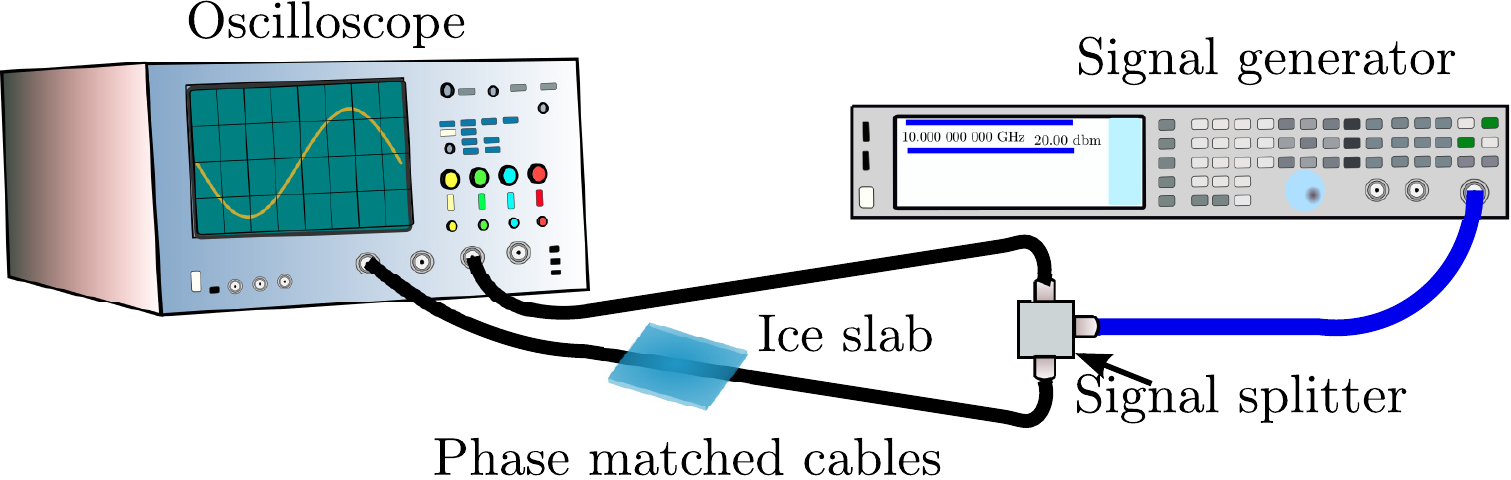}
\caption{Schematics of the setup used to record microwave signals. The signal from the generator is split into two parts using a calibrated splitter. The identical signals from the splitter are fed into the input channels of the oscilloscope using phase matched cables. The signals are digitized at the sampling rate of 80 GS/s and stored in the internal memory of the oscilloscope. The temperature of one of the phase matched cables  is varied by using ice slabs to measure the phase difference between the signals due to temperature differences.}
\label{SETUP}
\vspace{-0.25in}
\end{figure}

\textbf{Setup for the data acquisition.} The schematics of the setup used to digitally record the microwave signals is shown in Fig.\ref{SETUP}. In the measurements the signal from signal generator (N5183A MXG microwave analog signal generator) is split into two identical replicas using a calibrated signal splitter. The two signals are fed into channel 1 and 3 of the oscilloscope (DSAX93204A Infiniium High-Performance Oscilloscope) using phase matched cables (15443A Matched cable pair, 90 cm long at room temperature).\footnote{ All the equipment used in the experiment are provided by Agilent Technologies.} Signals at frequencies between 10 - 30 GHz are used in the measurements. The sampling rate of signal digitization is set to 80 GS/s per channel. The digitized signals are first stored in the internal memory (2 GB per channel) of the oscilloscope. After each recording, the data are transferred to external storage medium (computer hard disks). Signal analysis is done on the stored data using digital cavity and generalized lock-in techniques.   

\section{Signal analysis.} Digital cavities with $N_c= 10^6 - 2\times 10^8$ are used to select the desired frequency component from the microwave signal. The resonant frequency of the cavity is set to 16 GHz ($n=5,\quad \Delta t=12.5$ ps). Up-shifting of the signal by multiplying with appropriate cosine functions is done to scan the contributions of different frequency components in the signal close to the expected main frequency component. The phase and amplitude of the waveform that builds up in the cavity are analyzed to verify the algorithms for precision measurement of microwave signals.

\section{Results and discussion}
\subsection{Precision measurement of microwave signals} 
\begin{figure}[!t]
\centering
\includegraphics[width=3.5in]{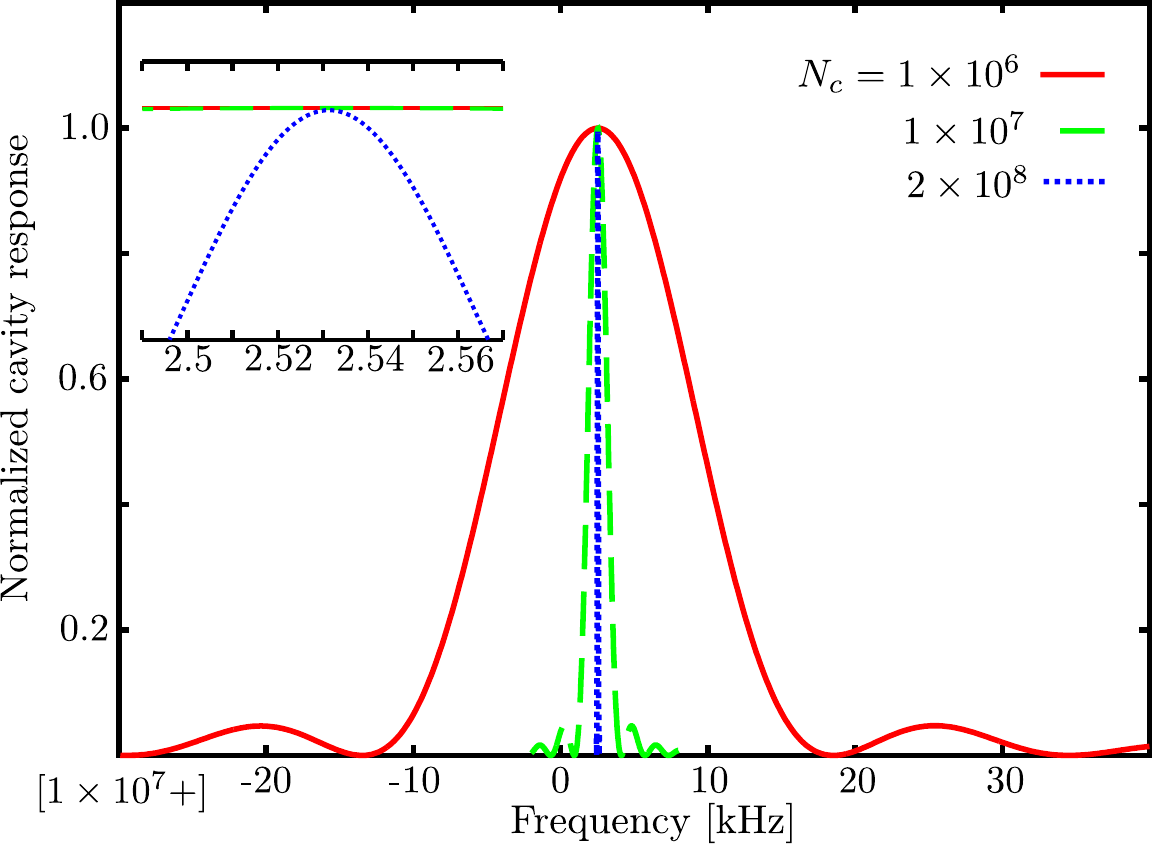}
\caption{Response of the digital cavity to the signal around 10 GHz for different values of $N_c$. The maximum of the response is at $(1\times 10^7+2.53)$ kHz and the FWHM is about 70 Hz for $N_c= 2\times 10^8$ (shown in the inset).}
\label{PMS}
\vspace{-0.25in}
\end{figure}

The response of the digital cavity to signals around 10 GHz  is shown in Fig.\ref{PMS}. The FWHM (full width at half maximum) of the signal around the maximum narrows progressively for the different values of the cavity fold $N_c$. The FWHM for $N_c=2\times 10^8$ shown in the inset is about 70 Hz. The maximum of the cavity response is at $(1\times 10^7+2.53)$ kHz, which is  2.53 kHz more than the frequency value set in the signal generator. The apparent difference between the frequency set in the signal generator and the oscilloscope gives a measure of the difference in the clock speeds. In our setup the clock in the oscilloscope lags by  about 0.25 $\mu$s per every second of the time lapsed in the signal generator. 

The timing precision in the measurements can also be calculated by measuring the average relative phase ($\Delta\phi$), i.e. the phase difference, between the identical signals acquired by the two channels as shown in Fig.\ref{SETUP}. The relative phase at the room temperature, after correcting for small drifts, for the signal at 10 GHz is 0.350$\pm$ 0.02 degrees. The uncertainty in the measurement is the standard deviation computed from 10 independent measurements carried out within 10 minutes. The relative phase of 0.35 degrees corresponds to a time difference of 0.1 ps, which is less than the timing tolerance of 25 ps specified for the phase matched cable.
The uncertainty of 0.02 degrees corresponds to an uncertainty of $\approx 6$ fs in the measurement of time.\footnote{The tolerance value specified by the supplier is an upper limit. Our measurements provide accurate values.} 

The ability to measure microwave signals with a resolution of few tens of hertz is very attractive to improving noise rejection and precision in a wide variety of applications such as in radar/sonar systems,~\cite{SKOLNIK_2001, CIMINI_2008,COLEMAN_2008} clock synchronization,~\cite{SERPEDIN_2011, CHENG_2011} remote sensing,~\cite{SIEGEL_2008,COLEMAN_2008} etc. Apart from such applications, extreme precision in the measurement of phase and amplitude of short wavelength electromagnetic (EM) waves allows us to use the technique to measure other physical quantities with high precision. In the following sub-sections we use the technique to study the effect of cooling on the cable carrying the signal.

\subsection{Newton's law of cooling by the precision measurement of the phase change.} The setup used in the measurements is the same as described previously (Fig.\ref{SETUP}). Apart from measuring the signals at the room temperature, one of the phase matched cables is cooled by using ice slabs. The data is recorded for 25 ms at different intervals. As the temperature decreases, the cable contracts, which leads to the change in the relative phase of the signal between the two cables. The change in the relative phase, $\Delta \phi(t)$ as the cooling time $t$  progresses is given by (see Appendix  for the derivation):
\begin{equation}\label{EQ3}
\Delta \phi(t) = X_0+X_1 \exp(-k t);
\end{equation}
where $k$ is the rate of cooling. We use $X_0$ and $X_1$ as fitting parameters. 

\begin{figure}[!t]
\centering
\includegraphics[width=3.2in]{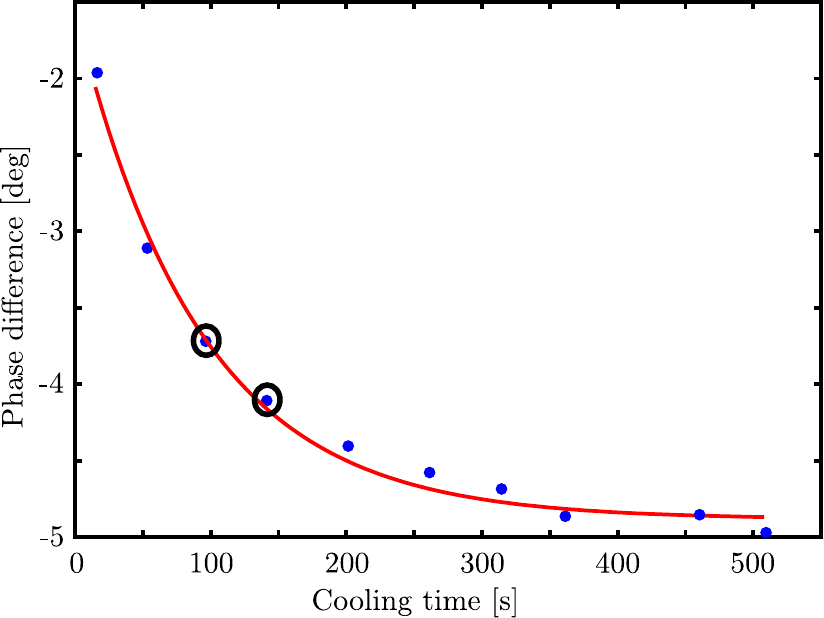}
\caption{The change in the difference in the phase as a function of cooling time. The decrease in the phase difference follows an exponential decay due to the Newton's law of cooling.}
\label{COOLING}
%\vspace{-0.25in}
\end{figure}

The change in the relative phase at different time points during the cooling is shown by the points in Fig.\ref{COOLING}. In accordance to \eref{EQ3} the data points follow an exponential decrease of the relative phase with respect to time. The red curve is the fit of \eref{EQ3} to the data points. The parameters obtained from the fit are $k= 0.01\pm 0.001$ s$^{-1}$ and $X_1 = 3.3\pm 0.1$ deg. As the change in the relative phase is proportional to the change in the length of the cable (Appendix , \eref{APB_EQ2}), $k$ also gives the time constant for the contraction of the cable. The contraction in the length between the time points shown by circles in the Fig.\ref{COOLING} is about 32 $\mu$m.  Note that the value we have calculated depends on the choice of wavelength $\lambda_c$. We have used $\lambda_c= 29.979246$ mm for 10 GHz signal, which is equal to the wavelength of the corresponding EM wave in the vaccuum. This value may be lower for the signals travelling in a cable. Hence, the change in the length we have calculated gives an upper limit rather than the exact value. Nevertheless, the calculation we have presented serves as a proof of principle measurment of the length change. Similarly, the upper limit to the precision with which we can measure the contraction can be calculated from the measurement precision of the relative phase. The calculated value is 1.7$\mu$m, which corresponds to the precision in the phase measurement of 0.02 degrees. Using \eref{APB_EQ3} with the value of the thermal expansion coefficient, $\alpha = 16.5\times 10^{-6}$ K$^{-1}$ (for copper), we get a precision of 0.11 $^{\circ}$C in the measurement of the change in the temperature. Alternatively, in a controlled experiment where the temperature change is measured externally, the technique can be used to measure the speed of the EM signal in the cable and therefore the wavelength with extreme precision. Note that the methodology we have presented does not only allow for the precision measurement of the different physical quantities but also that these measurements can be done within few tens of milliseconds with real time signal analysis.

\subsection{Newton's law of cooling by the precision measurement of the amplitude.}  
\begin{figure}[!t]
\centering
\includegraphics[width=3.5in]{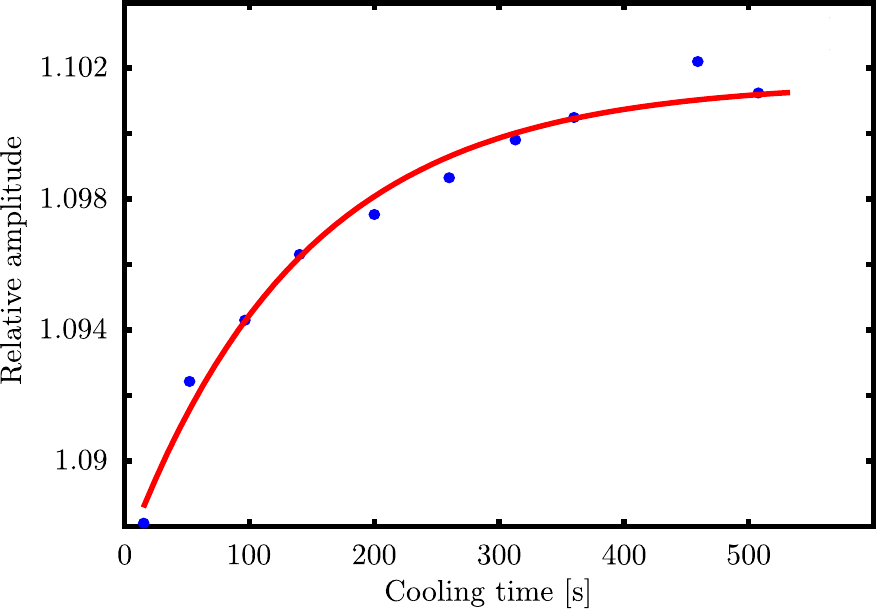}
\caption{The change in the relative amplitude at different times during the cooling of one of the cables. The relative amplitude of the signal grows asymptotically with time showing decrease in the resistance of the cable with time.}
\label{COOLINGB}
\vspace{-0.25in}
\end{figure}

The amplitudes of the signals described previously provide further information about other physical properties of the cables. Here, we focus on the change in the amplitude of the signal caused by the change in the resistance due to the change in the temperature. For most of the metallic wires the resistance decreases with the decrease in temperature, and this leads to the increase in amplitude, $A$, of the signal recorded by an oscilloscope. The variation of the relative amplitude during the cooling is given by (see Appendix  for derivation):
\begin{equation}\label{EQ4}
A(t) = \frac{C_0}{1+r_0[1-0.136(1-\exp(-k t))]},
\end{equation}
 where $C_0$ is a constant. We have used $\beta = 0.0068$ K$^{-1}$ (temperature coefficient of resistivity) and $T_0-T_a = 20$ K in \eref{APC_EQ9} ($T_0$ is the room temperature and $T_a$ is the temperature of ice). 
  Fig.\ref{COOLINGB} shows the change in the relative amplitude during the cooling. The red curve in Fig.\ref{COOLINGB} is \eref{EQ4} fitted to the data points. The rate of change in the relative amplitude obtained from the fit is $k=0.008\pm0.001$ s$^{-1}$. The rate of change calculated from the measurement of the amplitude differs by about 20\% from the rate of change calculated from the measurement of the phase change. Considering the linearity we have assumed in the thermal expansion as well as the electrical resistivity this discrepancy is reasonable. Moreover, the measurements are affected by the external sources of noise. Precision measurement of temperature effects on thermal expansion as well as resistivity requires extreme control of the environment.\cite{THERMAL_EXPANSION, THERMAL_PROPERTIES}

 \section{Limitation and scope of the technique.} 
The measurements we have done have not been calibrated, thus they have not been used to compute the physical properties like speed of signal in the cable or resistivity of the cable. Yet the measurements show that such quantities, in principle, can be directly calculated by measuring the phase and amplitude of the signals passing through the materials. Better control of the environment can improve the precision in such measurements. The limitation in the precision of our measurements is due to digitization noise, noise in the signal generator and fluctuations in the environment. If only the digitization error is considered then the uncertainty in the phase measurement using an 8 bit digitizer is at most 10$^{\circ}$ per cycle of the signal. Averaging over $10^8$ cycles reduces the uncertainty to 0.001$^{\circ}$. Clearly the uncertainty in the phase (0.02$^{\circ}$) in our measurements is affected by sources other than the digitizer itself. More can be done to reduce the noise from the source of the signal and environmental fluctuations. 

Modern digitizers with the highest bandwidth can allow measurement of signals up to 65 GHz. As the high speed digitizers are being continuously improved, measurement of even higher frequency signals would be possible in the future.  One of the areas where extreme precision measurement of microwaves is likely to have scientific importance is in accelerators that use microwaves to accelerate particles to relativistic velocities.\cite{VALUCH_2009} Apart from these well known applications we have shown with examples that precision measurement of high frequency signals through material allow us to monitor the physical changes in the material, which makes the technique very attractive in virtual metrology.~\cite{KAO_2007} Real time implementation of generalized lock-in amplifier for measuring microwave signals would enable accurate real-time measurements for virtual metrology.~\cite{WU_2008}

\section{Conclusion}
We have shown that generalized lock-in amplifier can be used to measure giga-hertz signals with a precision of few tens of hertz. Our measurements show that the change in the phase of a signal at 10 GHz can be measured with a precision of 0.02$^{\circ}$. Such precise measurements allow us to directly measure the physical changes in the medium through which the signal is propagating. These changes can be detected within a fraction of a second, which makes the technique promising for real time virtual metrology.

\section*{Acknowledgment}

The authors thank Agilent Technologies and Testhouse Nordic for providing the signal generator and oscilloscope for the measurements. This work would not have been possible without the technical support of Thomas G\"oransson from Agilent Technologies.  Financial support from the Knut and Alice Wallenberg (KAW) Foundation, the Wernner-Gren Foundation, the Swedish Foundation for International Cooperation in Research and Higher Education (STINT), Swedish Research Council, and Lund University Innovation System  is gratefully acknowledged.

\appendix

\section{Derivation of the relation between the relative phase and the cooling time}

The cable contracts when it is cooled. If the change in the temperature, $\Delta T$ is small, the contraction in length of the cable,  $\Delta L$, is given by~\cite{THERMAL_PROPERTIES}
\begin{equation}\label{APB_EQ1}
\Delta L = -\alpha \Delta T L,
\end{equation}
where $\alpha$ is the linear expansion coefficient and $L$ is the length of the cable.

The relative phase ($\Delta \phi$), i.e. the difference in the phase between the signals through the two cables, changes with the contraction of the wire. It is given by
\begin{equation}\label{APB_EQ2}
\Delta \phi = \frac{360^{\circ} }{\lambda_c}\Delta L + O,
\end{equation}
where $\lambda_c$ is the wavelength of the EM waves in the cable and the phase is measured in degrees and $O$ is the offset. 

From \eref{APB_EQ1} and \eref{APB_EQ2}, we see that the change in the temperature leads to the change in the realtive phase:
\begin{equation}\label{APB_EQ3}
\Delta \phi = -\frac{360^{\circ}}{\lambda_c}\alpha \Delta T L + O.
\end{equation}

The temperature, $T(t)$, of the cable as a function of time, $t$, when it is cooling down is given by the Newton's law of cooling:~\cite{APPL_CALCULUS}
\begin{equation}\label{APB_EQ4}
T(t)=T_a+(T_0-T_a)\exp(-k t),
\end{equation}
where $T_a$ is the temperature of ice (0 $^0$C), $T_0$ is the room temperature (21 $^0$C) and $k$ is the cooling constant. The change in the relative phase as a function of time is then given by
\begin{eqnarray}\label{APB_EQ5}
\Delta \phi(t) &=&-\frac{360^{\circ}}{\lambda_c}\alpha L\left(T_0-T(t)\right)\nonumber\\
               &=&  X_0+X_1\exp(-kt),
\end{eqnarray}
where $X_1= 360^{\circ} \alpha L (T_0-T_a)/\lambda_c$ is a constant in the measurements.

\section{Derivation of the relation between the relative amplitude and the cooling time}
The circuit diagram equivalent to the setup is shown in Fig.%\ref{CIRCREF}.
  As the effect of inductance is minor compared to the effect of temperature change, we use the following circuit equation:
\begin{equation}\label{APC_EQ1}
I(t) = \frac{V(t)}{R_0 + R(t)},
\end{equation}
where $V(t)$ is the voltage of the AC power source (the signal generator), $I(t)$ is the current, $R_0$ is the impedance in the receiver of the oscilloscope and $R(t)$ is the resistance of the cable. The voltage drop $V_0$ across the resistance $R_0$ is 
\begin{equation}\label{APC_EQ2}
V_0(t) = V(t)-I(t) R(t).
\end{equation}

Resistance of the cable, $R(t)$ at a particular temperature is given by:
\begin{equation}\label{APC_EQ3}
R(t) = \frac{\rho(T(t))\cdot L(T(t))}{S_0},
\end{equation}
where $\rho$ is the electrical resistivity of the cable and $S_0$ is the cross section of the cable. The length as well the resistivity  depend on the temperature. As the temperature difference is not large, we use the linear approximation in the thermal expansion coefficient $\alpha$ and the temperature coefficient of resistivity, $\beta$:
\begin{equation}\label{APC_EQ4}
L(T)=L_0[1+\alpha(T-T_0)]
\end{equation}
and
\begin{equation}\label{APC_EQ5}
\rho(T)=\rho_0[1+\beta(T-T_0)].
\end{equation}
During the cooling, $T(t)$ is given by Newton's law of cooling (\Eref{APB_EQ4}). Using \eref{APB_EQ4},\eref{APC_EQ3},\eref{APC_EQ4} and \eref{APC_EQ5} the resistance of the cable during the cooling can be written as a function of time:
\begin{eqnarray}\label{APC_EQ6}
R(t) &=&\frac{\rho_0 L_0}{S_0} [1-(\alpha + \beta)(T_0-T_a)(1-\exp(-kt))\nonumber\\
     &&+\alpha\beta(T_0-T_a)^2(1-\exp(-kt))^2]\nonumber\\
&\simeq&\frac{\rho_0 L_0}{S_0}[1-(\alpha + \beta)\nonumber\\
&&(T_0-T_a)(1-\exp(-kt))].
\end{eqnarray}
As $\alpha \ll \beta$ for copper, we can further simplify \eref{APC_EQ6} to 
\begin{equation}\label{APC_EQ7}
R(t) = \frac{\rho_0 L_0}{S_0} [1-\beta(T_0-T_a)(1-\exp(-kt))].
\end{equation} 

Using \eref{APC_EQ1}, \eref{APC_EQ2} and \eref{APC_EQ7} we can get the expression for the voltage read by the oscilloscope:
\begin{equation}\label{APC_EQ8}
V_0(t) = V(t)\cdot \frac{1}{1+r_0[1-\beta(T_0-T_a)(1-\exp(-kt))]},
\end{equation}
where $r_0=(\rho_0 L_0)/(R_0S_0)$.

Finally, we take the ratio of the voltage reading of the signal through the cooled cable to that of the voltage reading of the cable at the room temperature to reduce the errors in the measurement caused by the fluctuations in the signal generator as well as the environment.
\begin{equation}\label{APC_EQ9}
A(t) = \frac{C_0}{1+r_0[1-\beta(T_0-T_a)(1-\exp(-kt))]},
\end{equation}
where $C_0$ is a constant.

\section*{References}
%\bibliographystyle{unsrt}
%\bibliography{IEEEabrv,../../../masterb}

\end{document}